\documentclass[iop]{emulateapj}

\newcommand{\etal}{et al.}

\newcommand\chandra{{\it Chandra}}
\newcommand\xmm{{\it XMM-Newton\/}}
\newcommand\ro{{\it ROSAT\/}}
\newcommand\fermi{{\it Fermi}}
\newcommand\calv{1RXS J141256.0$+$792204}
\newcommand\psr{PSR J1740+1000}
\newcommand\gr{$\gamma$-ray}
\newcommand\grs{$\gamma$-rays}

\def\simlt{\mathrel{\hbox{\rlap{\hbox{\lower4pt\hbox{$\sim$}}}\hbox{$<$}}}}
\def\simgt{\mathrel{\hbox{\rlap{\hbox{\lower4pt\hbox{$\sim$}}}\hbox{$>$}}}}

\slugcomment{}

\received{2013 August 7}
\accepted{2013 October 21}

\shorttitle{Spin-down of Calvera}
\shortauthors{Halpern, Bogdanov, Gotthelf}

\begin{document}

\title{X-ray Measurement of the Spin-Down of Calvera:\\
a Radio- and Gamma-ray-Quiet Pulsar}

\author{J. P. Halpern, S. Bogdanov, and E. V. Gotthelf}
\affil{Columbia Astrophysics Laboratory, Columbia University,
550 West 120th Street, New York, NY 10027-6601, USA;
jules@astro.columbia.edu}

\begin{abstract}
We measure spin-down of the 59~ms X-ray pulsar Calvera by comparing
the \xmm\ discovery data from 2009 with new \chandra\ timing
observations taken in 2013.  Its period derivative is
$\dot P=(3.19\pm0.08)\times10^{-15}$, which corresponds
to spin-down luminosity $\dot E=6.1\times10^{35}$ erg~s$^{-1}$,
characteristic age $\tau_c\equiv P/2\dot P=2.9\times10^5$~yr, and surface
dipole magnetic field strength $B_s=4.4\times10^{11}$~G.
These values rule out a mildly recycled pulsar, but Calvera
could be an orphaned central compact object (anti-magnetar),
with a magnetic field that was initially buried by supernova
debris and is now reemerging and approaching normal strength.
We also performed unsuccessful searches for high-energy \grs\
from Calvera in both imaging and timing of $>100$~MeV \fermi\ photons. 
Even though the distance to Calvera is uncertain by an order of magnitude,
an upper limit of $d<2$~kpc inferred from X-ray spectra
implies a \gr\ luminosity limit of $<3.3\times10^{32}$ erg~s$^{-1}$,
which is less than that of any pulsar of comparable $\dot E$.
Calvera shares some properties with \psr, a young radio pulsar
that we show by virtue of its lack of proper motion was born
outside of the Galactic disk.
As an energetic, high-Galactic-latitude pulsar,
Calvera is unique in being undetected in both radio and \grs\
to faint limits, which should place interesting constraints
on models for particle acceleration and beam patterns
in pulsar magnetospheres.
\end{abstract}


\section{Introduction}

The neutron star (NS) candidate \calv, dubbed ``Calvera,'' by
\citet{rut08}, was selected from the \ro\ All-Sky Survey and
observed by \chandra\ \citep{rut08,she09}.  It was not until
a pair of \xmm\ observations was obtained with high time resolution
that \citet{zan11} discovered the 59~ms pulsations from Calvera.
The X-ray emission from Calvera is best described by a two-temperature
(blackbody or hydrogen atmosphere) spectrum, with $kT$
in the range 0.1--0.25~keV \citep{she09,zan11}.  
The fitted column density in these models is equal to or greater
than the Galactic value of $N_{\rm H}=2.7\times10^{20}$~cm$^{-2}$
\citep{kal05} which, at its high Galactic latitude of $+37^{\circ}$,
leaves Calvera's distance highly uncertain.
Its bolometric luminosity is $L_X\approx1.2\times10^{31}\ d_{300}^2$
erg~s$^{-1}$, where $d_{300}$ is distance in units of 300~pc.  

Calvera remains radio quiet even after
deep searches for radio pulsations at the known period \citep{hes07,zan11}.
Analyzing data from the \fermi\ Large Area Telescope (LAT), \citet{zan11}
concluded that \gr\ pulsations are detected from Calvera at
$>100$~MeV. \citet{hal11} reanalyzed the \xmm\ timing data
in detail together with additional \fermi\ data, showing
that the claimed \fermi\ detection was not real.

\begin{figure}[t]
\begin{center}
\includegraphics[width=0.99\linewidth,angle=0]{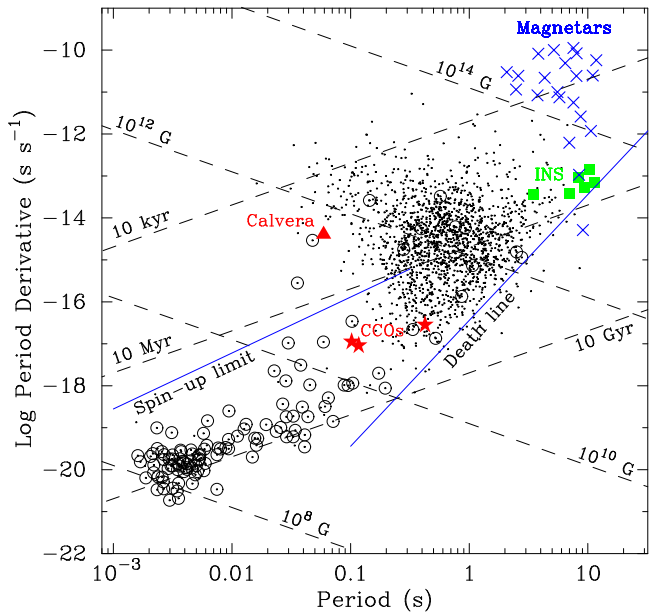}
\end{center}
\vspace{-0.1in}
\caption{
Pulsar populations on the $(P$,$\dot P)$ diagram,
including INSs (green squares), magnetars (blue crosses),
CCOs (red stars), and Calvera (red triangle).  Black dots are
isolated pulsars and circled dots are pulsars in binaries.
(Pulsars in globular clusters are excluded as their period
derivatives are not entirely intrinsic.)
The \citet{van87} spin-up limit for recycled pulsars corresponds
to $P({\rm ms})=1.9\,(B/10^9\,{\rm G})^{6/7}$.
The radio pulsar death line $B/P^2=1.7\times10^{11}$~G~s$^{-2}$ of
\citet{bha92} is indicated.
(A color version of this figure is available in the online journal.)
}
\label{fig:ppdot}
\end{figure}

Calvera is of interest largely because its classification
among the families of NSs is unclear, and it may occupy
a unique role.   Its properties distinguish it from
the seven other isolated NSs (INSs; \citealt{hab07}) that were
discovered by \ro, which are slowly rotating ($P=3-11$~s), cooler
NSs in the solar neighborhood (Figure~\ref{fig:ppdot}).  X-ray timing and
spectroscopy, and kinematic studies of the INSs indicates that they
have strong magnetic fields, $B_s\approx2\times10^{13}$~G, and are
$\approx 10^6$~yr old \citep{kap09}.  Calvera has at least twice the
temperature of the INSs, but its age and magnetic field strength were
undetermined until now.  These unknowns allowed \citet{zan11} and \citet{hal11}
to hypothesize that Calvera is either a mildly recycled pulsar
\citep[as defined by][]{bel10}, or the first candidate for the
elusive descendants
of the central compact objects (CCOs) in supernova remnants, thermally 
emitting NSs that have weak dipole fields in the range
$(0.3-1)\times10^{11}$~G \citep{got13}.  However, if $B>10^{11}$~G,
it could just be an ordinary rotation-powered pulsar.
In this Paper, we report the detection of Calvera'a spin-down, which 
reveals its characteristic age, spin-down power, and dipole magnetic field
strength.  We also derive a new, interesting upper limit on its \gr\
luminosity.

%-----------------------------Figure Start------------------------------
%-----------------------------Figure End--------------------------------

\begin{deluxetable*}{lcccccccr}
\tabletypesize{\footnotesize}
\tablewidth{0pt}
\tablecaption{Timing of Calvera}
\tablehead{
\colhead{Mission} & \colhead{Instr/Mode} & \colhead{ObsID} &
\colhead{Date (UT)} & \colhead{Date (MJD)} & \colhead{Exp. (s)} &
\colhead{Photons\tablenotemark{a}} &
\colhead{Frequency (Hz)\tablenotemark{b}} & \colhead{$Z_1^2$}
}
\startdata
\xmm\     & EPIC-pn/SW & 0601180101 & 2009 Aug 31  & 55,074.30 & 13,941 &  7703  & 16.8924052(25) & 141.1 \\
\xmm\     & EPIC-pn/SW & 0601180201 & 2009 Oct 10  & 55,114.18 & 19,477 & 10,515 & 16.8924041(15) & 201.9 \\
\chandra\ & ACIS-S3/CC & 13788      & 2013 Feb 12  & 56,335.81 & 19,679 &  2748  & 16.8923057(32) &  94.2 \\
\chandra\ & ACIS-S3/CC & 15613      & 2013 Feb 18  & 56,341.12 & 17,093 &  2463  & 16.8923083(30) & 117.8 
\enddata
\tablenotetext{a}{Background subtracted counts in the
0.15--2 keV band for \xmm, and the 0.2--4~keV band for \chandra.}
\tablenotetext{b}{1 sigma error in parenthesis.}
\label{tab:timing}
\end{deluxetable*}

\section{X-ray Observations}

A new observation of Calvera was obtained using the \chandra\
Advanced Camera for Imaging and Spectroscopy \citep[ACIS;][]{gar03}
operated in continuous-clocking (CC) mode to provide a time resolution
of 2.85~ms. To achieve fast timing in CC mode, one spatial
dimension of the CCD image (the row number) is lost.
ACIS has $0.\!^{\prime\prime}5$ pixels, comparable to the on-axis
point-spread function.  The target was placed on the back-illuminated
ACIS-S3 CCD.  Due to scheduling
restrictions, the observation was split into two parts separated
by five days in 2013 February, as listed in Table~\ref{tab:timing}.  
All data reduction and analysis was performed
with the \chandra\ Interactive Analysis of Observation software
\citep[CIAO;][]{fru06}
version 4.5, using the calibration database (CALDB) v4.1.3.
We extracted photons from five columns around the pulsar position, and
transformed their arrival times to the solar system barycenter
in Barycentric Dynamical Time (TDB) using the \chandra\ measured coordinates
given in \citet{she09}, R.A.=$14^{\rm h}12^{\rm m}55.\!^{\rm s}84$,
decl.=$+79^{\circ}22^{\prime}03.\!^{\prime\prime}7$ (J2000.0).
In order to subtract background from the pulse profile and spectrum,
background in CC mode was extracted from a region adjacent to the source.

%-----------------------------Figure Start------------------------------
\begin{figure}[t]
\begin{center}
\includegraphics[width=0.98\linewidth,angle=0]{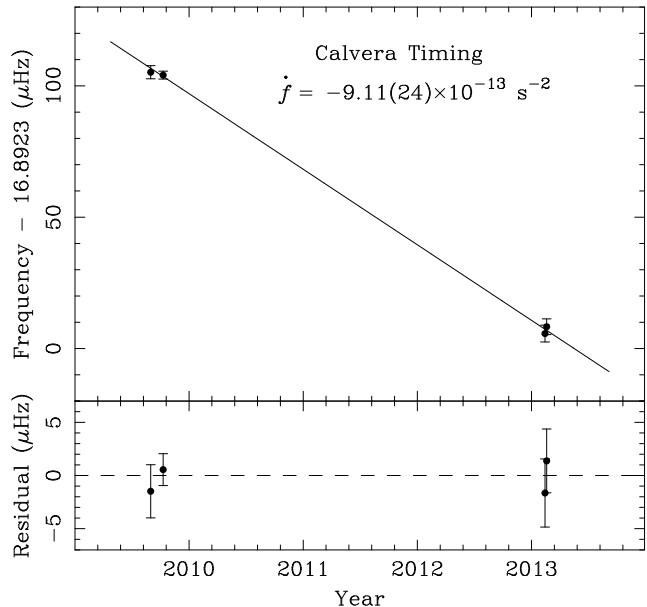}
\end{center}
\vspace{-0.1in}
\caption{
X-ray timing of Calvera using the measurements listed in Table~\ref{tab:timing}.
The fitted frequency derivative corresponds to
$\dot P=(3.19\pm0.08)\times10^{-15}$.  Residuals from the linear
fit are shown in the lower panel.
}
\label{fig:timing}
\end{figure}
%-----------------------------Figure End--------------------------------

The best fitting frequency was determined for each \chandra\
data set using the $Z^2_1$, or Rayleigh test \citep{str80,buc83}
on photons in the 0.2--4~keV range.
Frequency measurements from the two \xmm\ observations
of Calvera acquired by \citet{zan11}, as reduced by \citet{hal11},
are also listed in Table~\ref{tab:timing}, and shown in
Figure~\ref{fig:timing}. None of the observations are closely spaced
enough to link coherently and obtain more precise timing parameters.

The close agreement in period between the observations
separated by 40 and 5 days, and the lack of an optical counterpart
to faint limits \citep[$g>26.3$;][]{rut08}, tend to rule out
known binary scenarios.  The ten known double NS systems,
which have orbital periods of $0.1-19$ days, produce period changes
orders of magnitude larger than our uncertainties.
A wide binary with a very cool white dwarf and
period of order years could possibly accommodate the timing data
and optical limit if the distance is large, but such systems are rare.
A substellar companion would not be massive enough to account for
the period change.

In the absence of clear evidence for orbital motion, we interpret
the long-term decrease in frequency as intrinsic to an
isolated NS.  Accordingly, we determined the frequency
derivative with a $\chi^2$ fit to the four frequency measurements,
with the result that $\dot f=(-9.11\pm0.24)\times10^{-13}$~s$^{-2}$.
This value is only a factor of $\approx2$
smaller than the upper limit determined previously
from the two \xmm\ observations alone \citep{hal11}.
The corresponding spin-down properties are
$\dot E=6.1\times10^{35}$ erg~s$^{-1}$, $\tau_c=2.9\times10^5$~yr, and
$B_s=4.4\times10^{11}$~G.

%-----------------------------Figure Start------------------------------
\begin{figure}
\includegraphics[width=0.99\linewidth,angle=0]{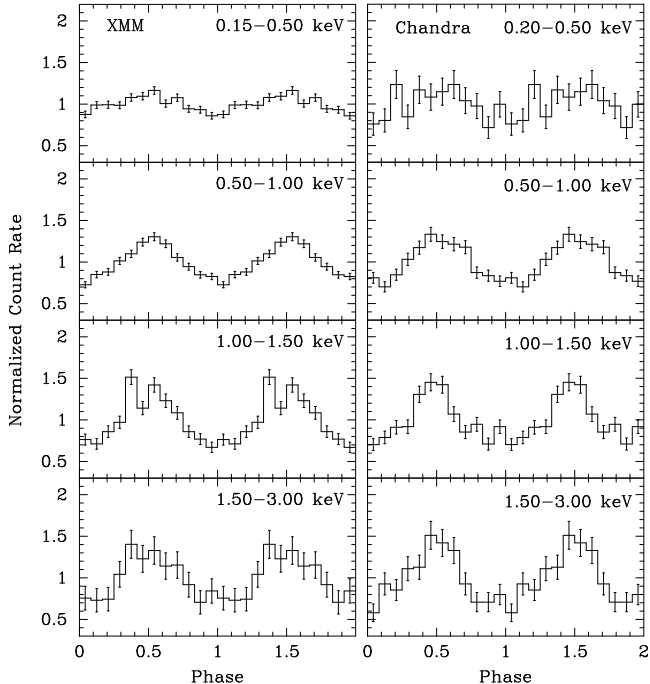}
\caption{
Normalized and background subtracted X-ray pulse profiles
of Calvera from the combined \xmm\ (left) and \chandra\ (right)
observations listed in Table~\ref{tab:timing}.  The TDB epoch of
phase zero is MJD 55094.0000004 for \xmm, and
MJD 56335.8100000 for \chandra.
}
\label{fig:pulse}
\end{figure}
%-----------------------------Figure End--------------------------------

The normalized and background subtracted pulse profiles
of Calvera are shown in Figure~\ref{fig:pulse} for both
\xmm\ and \chandra\ data.  We note the increase in pulsed
fraction as a function of X-ray energy.  This is especially clear
from \xmm, which has much higher throughput at the lowest energies
than \chandra.

Spectral fits to the combined Chandra CC-mode data yield results
similar to previous observations \citep{she09,zan11}.  Parameters
for a power-law plus blackbody fit, and a two-blackbody model,
are listed in Table~\ref{tab:spec}.  Even though the
power-law plus blackbody is a good fit, \xmm\ spectra
that have better statistics at low energy find that the
same model requires $N_{\rm H}$ to exceed the Galactic value of
$2.7\times10^{20}$~cm$^{-2}$ by a factor of three \citep{zan11}.
Therefore, we prefer the purely thermal model, which does not
suffer from this problem, but does require two temperatures,
i.e., the same number of parameters as the power-law plus blackbody.

\begin{deluxetable}{lc}
\tablewidth{0pt}
\tablecaption{Spectral Fits for ObsIDs 13788 and 15613 Combined}
\tablehead{
\colhead{Parameter}  & \colhead{Value\tablenotemark{a}}
}
\startdata
Exp. time (s)                            & 36,772 \\
Counts (s$^{-1}$)                         &  0.138  \\
\tableline
\hspace{0.6 in}Power-law plus blackbody model\\
\tableline
$N_{\rm H}$ (cm$^{-2}$)                       & $3.7^{+5.8}_{-3.7}\times10^{20}$ \\
$\Gamma$                                    & $3.7^{+0.6}_{-0.5}$             \\
$kT$ (keV)                                  & $0.259^{+0.012}_{-0.013}$        \\
$R$ (km)\tablenotemark{b}                   & 0.096                         \\
$F_x\ (0.3-10\ {\rm keV}$)\tablenotemark{c} & $1.09\times10^{-12}$           \\
$L_x\ (0.3-10\ {\rm keV}$)\tablenotemark{d} & $1.63\times10^{31}$            \\
$\chi^2_{\nu}(\nu)$                          & 1.004(70)                     \\
\tableline
\hspace{0.6 in} Two-blackbody model\\
\tableline
$N_{\rm H}$ (cm$^{-2}$)                       & $<3\times10^{20}$ \\
$kT_1$ (keV)                                & $0.090\pm0.006$               \\
$R_1$ (km)\tablenotemark{b}                 & 0.99                          \\
$kT_2$ (keV)                                & $0.264\pm 0.005$              \\
$R_2$ (km)\tablenotemark{b}                 & 0.11                          \\
$F_x\ (0.3-10\ {\rm keV}$)\tablenotemark{c} & $1.07\times10^{-12}$           \\
$L_x\ (0.3-10\ {\rm keV}$)\tablenotemark{d} & $1.15\times10^{31}$            \\
$\chi^2_{\nu}(\nu)$                          & 1.153(70)               
\enddata
\tablenotetext{a}{Uncertainties are 90\% confidence for one interesting
parameter.}
\tablenotetext{b}{Assuming $d=300$~pc.}
\tablenotetext{c}{Absorbed flux in units of erg cm$^{-2}$ s$^{-1}$.}
\tablenotetext{d}{Unabsorbed luminosity in units of erg s$^{-1}$,
for $d=300$~pc.}
\label{tab:spec}
\end{deluxetable}

\section{Search for $\gamma$-rays}

Given its high spin-down power and likely proximity, it is
surprising that Calvera is not a bright \fermi\ pulsar.
We can place a conservative upper limit of
$F_{\gamma}<3\times10^{-12}$ erg~cm$^{-2}$~s$^{-1}$
on its $>100$~MeV flux simply from
its absence in the 2GFL catalog \citep{nol12}, as this flux
is comparable to the weakest catalogued sources in its
immediate vicinity.   This corresponds to a luminosity upper
limit of $L_{\gamma}<1.4\times10^{33}\,d_{2\,\rm kpc}^2$ erg~s$^{-1}$,
assumed isotropic. (The adopted upper limit on Calvera's distance,
2~kpc, will be justified in Section~4.1).
In comparison, detected \gr\ pulsars
of similar $\dot E$ typically have $L_{\gamma}=10^{34-35}$~erg~s$^{-1}$.
Next, we make a more sensitive test for \gr\ emission at the position
of Calvera by using the additional \fermi\ data accumulated since
the construction of the 2GFL catalog, and also by searching for
pulsations around the known $f$ and $\dot f$.

To search for \gr\ point source emission from Calvera, we
retrieved Pass~7 \fermi\ LAT event data between 2008 August 4
and 2013 July 1 and within 20$^{\circ}$ of the pulsar position. The
analysis was carried out using the \fermi\ Science
Tools\footnote{Available at
  \url{http://fermi.gsfc.nasa.gov/ssc/data/analysis/scitools/overview.html}}
v9r27p1. Following the recommended analysis guidelines from the \fermi\
Science Support Center, the data were filtered for ``source'' class
events in good time intervals with energies above 100 MeV, zenith
angles smaller than 100$^{\circ}$, and telescope rocking angles
$\le52^{\circ}$ using the {\tt gtselect} and {\tt gtmktime} tools.

%-----------------------------Figure Start------------------------------
\begin{figure*}
\begin{center}
\includegraphics[width=1.0\textwidth]{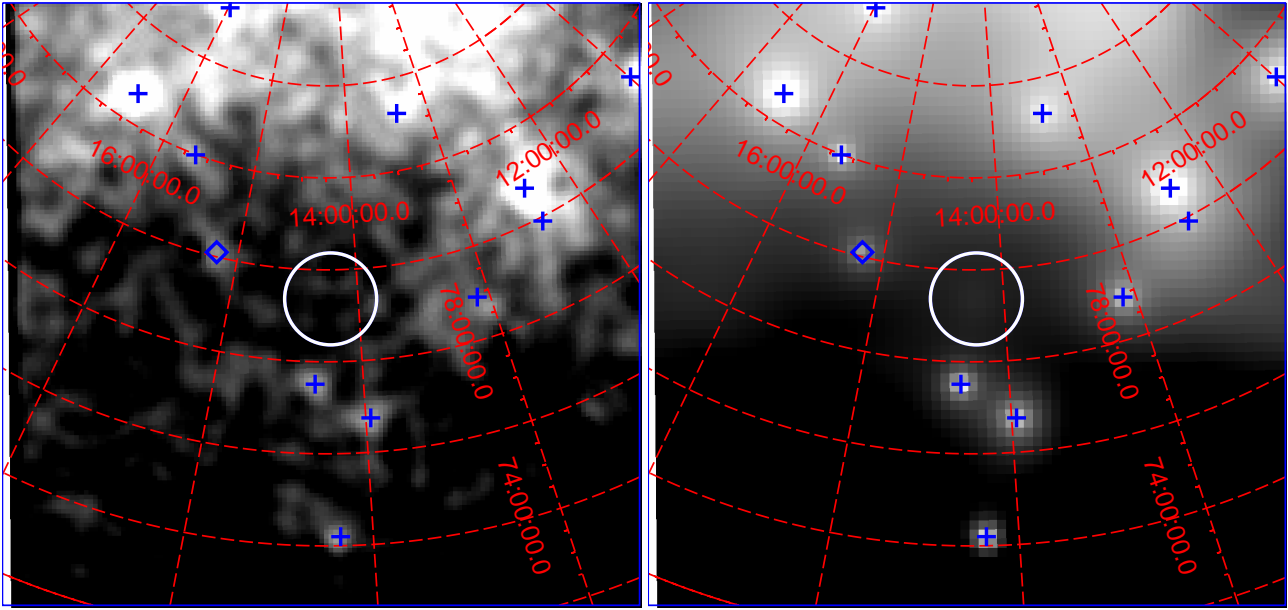}
\end{center}
\vspace{-0.1in}
\caption{\fermi\ LAT 0.1--300 GeV counts map (left) and likelihood
  analysis output model map (right). The blue crosses mark the
  positions of 2FGL catalog point sources, while the diamond marks the
  position of the source we added to the model, likely identified
  with CRATES J151032.75+800005.3.  The white circle
  of radius $1^{\circ}$ is centered on the X-ray position of Calvera.
  (A color version of this figure is available in the online journal.)}
\label{fig:FermiImg}
\end{figure*} 
%-----------------------------Figure End--------------------------------

It is evident from the resulting \fermi\ LAT counts map
(Figure~\ref{fig:FermiImg}, left) that there is no obvious \gr\
source positionally coincident with Calvera.  To determine formally
whether Calvera is a \gr\ point source, we carried out a binned
likelihood analysis with the {\tt gtlike} tool based on the input
counts, exposure, and source maps, livetime cube and source model
generated with the \fermi\ Science Tools.  The input spectral
model for the region of interest included contributions from the
putative \gr\ pulsar counterpart, all 2FGL catalog sources
within $15^{\circ}$ of the pulsar, the extragalactic diffuse emission,
and the residual instrumental background, jointly modeled using the
{\tt iso\_p7v6source} template, and Galactic diffuse emission, modeled
with the {\tt gal\_2yearp7v6\_v0} map cube.

The first iteration of the likelihood analysis
revealed excess emission $2.\!^{\circ}7$ ENE of
Calvera's position that does not appear to be associated with any 2FGL
source (see Figure~\ref{fig:FermiImg}). Hence, we added a model source
described by a power-law at its position, (J2000.0)
R.A.=$15^{\rm h}10^{\rm m}53.\!^{\rm s}6$,
decl.=$+80^{\circ}05^{\prime}50^{\prime\prime}$ (J2000.0).
The likelihood analysis assigned it a test statistic value
of 33.7, corresponding to a significance of $\approx 5.8\sigma$.
This source is $<6^{\prime}$ from a radio
source with an inverted spectrum at $0.3-8.5$~GHz, listed as
CRATES J151032.75+800005.3 by \citet{hea07}, which also has
an X-ray counterpart, 1RXS J151026.3+795946, and magnitudes
from the Wide-field Infrared Survey Explorer\footnote{Data at 
http://wise2.ipac.caltech.edu/docs/release/allsky/} \citep{wri10},
$W1=13.428(24), W2=12.384(23), W3=9.584(26), W4=7.317(67)$,
that are typical of \gr\ blazars \citep{mas12}.
These blazar properties give us added confidence
that the new \gr\ source is real.
Its redshift is unknown, although it has a faint optical counterpart
(on the Palomar Observatory Sky Survey) that could lead to a measurement.

As appropriate for
a pulsar, we model Calvera's \gr\ spectrum as a power-law with
an exponential cutoff, of the form $dN/dt\propto E^{-\Gamma}
\exp(-E/E_c)$, where $\Gamma$ is the photon spectral index and $E_c$
is the cutoff energy.  The spectral parameters of the
pulsar, the 16 sources within 10$^{\circ}$ and the
normalization factors of diffuse components were left free in the
fit. For sources between 10$^{\circ}$ and $15^{\circ}$ away
the spectral parameters were kept fixed.

\begin{figure}
\begin{center}
\includegraphics[width=1.\linewidth,angle=0]{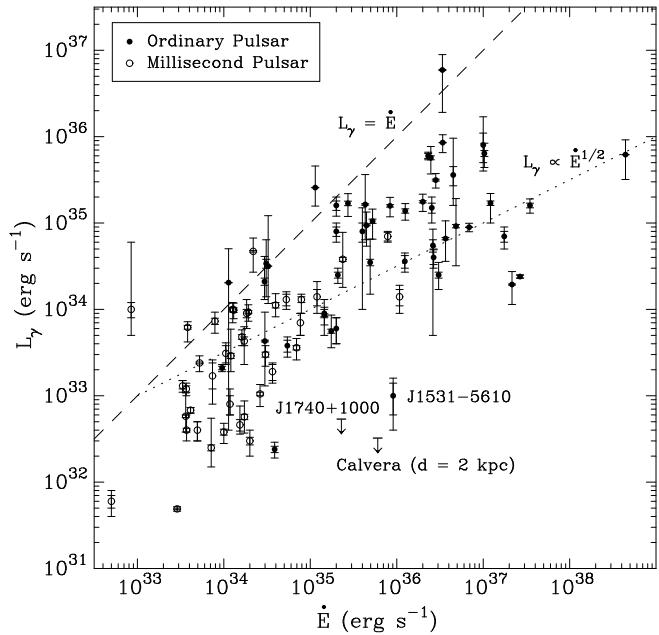}
\end{center}
\vspace{-0.1in}
\caption{
\gr\ luminosity (assumed isotropic)
of pulsars in the 0.1--100~GeV band
versus spn-down power, adapted from Figure~9 of the Second
\fermi\ Large Area Telescope Catalog of Gamma-Ray Pulsars
\citep{abd13}.  Two types of errors are shown for each point,
involving flux and distance.
Usually the uncertainty in distance gives a larger error bar
than the statistical uncertainty in flux.
The upper limit for \psr\ at its nominal distance of 1.24~kpc,
taken from \citep{rom11}, is also shown, as well as the upper
limit for Calvera for its maximum distance of 2~kpc.
The luminosity of PSR J1531$-$5610 represents its pulsed emission only.
}
\label{fig:fermi}
\end{figure}

The likelihood analysis gives a Test Statistic
\citep[TS; for a definition see][]{nol12}
value of 3.9 for Calvera, which corresponds to
only a $\sim$2$\sigma$ detection.  The spectral parameters are poorly
constrained.  The low statistical significance indicates that the
addition of a source at Calvera's position is not warranted by the
existing \fermi\ LAT data, implying that the pulsar is not a
\gr\ source.  Given the non-detection, to estimate the upper
limit on Calvera's \gr\ flux we followed the procedure used by
\citet{rom11}. In particular, we computed an expected photon index
and an exponential cutoff energy using the empirical relations
$\Gamma=-4.1+0.156\log_{10}\dot{E}$ and
$E_c\,({\rm GeV})=-0.45+0.71\log_{10}B_{\rm LC}$,
where $B_{\rm LC}\simeq9.3\,P^{-5/2}\,(\dot{P}/10^{-15})^{1/2}$~G 
is the magnetic field at the light cylinder.  This predicts
$\Gamma=1.5$ and $E_c=2.6$ GeV for Calvera. We then repeated the
likelihood analysis with Calvera's $\Gamma$ and $E_c$ fixed at these
values, while allowing the flux normalization to vary. Based on the
resulting value, the {\tt UpperLimit} Python module from the \fermi\
Science Tools gives $<8.2\times10^{-10}$ photons~cm$^{-2}$
s$^{-1}$, translating to an energy flux of $<6.8\times10^{-13}$
erg~cm$^{-2}$~s$^{-1}$ in the $0.1-300$ GeV interval.
The corresponding upper limit on luminosity,
$L_{\gamma}<3.3\times10^{32}\,d_{2\,\rm kpc}^2$ erg~s$^{-1}$,
is illustrated in Figure~\ref{fig:fermi} in comparison with detections
of 88 pulsars from the Second \fermi\ Large Area Telescope Catalog of
Gamma-Ray Pulsars \citep{abd13}.

%-----------------------------Figure Start------------------------------
%-----------------------------Figure End--------------------------------

Despite the lack of a \gr\ detection in the image plane, a search
for pulsed emission is potentially more sensitive, especially if
the signal is sharply peaked.  Accordingly, we searched for
\gr\ pulsations at the position of Calvera in aperture radii ranging from
$1.\!^{\circ}0-2.\!^{\circ}0$, in $0.\!^{\circ}5$ increments, for photons
in both the $0.1-300$~GeV and $0.3-300$~GeV ranges. 

Lacking a true ephemeris, we searched around the approximately
determined spin frequency $f$ and spin-down rate $\dot f$.
A large grid search is necessitated by the paucity of $\gamma$-ray
photons, which requires a time step of at least 100 days during
which phase coherence is assumed to be maintained.  The number of
trials was reduced by confining the range of parameters searched
to a set of five possible aliases allowed by the close pair of
\chandra\ observations.  In each case, conservative
$\pm3\sigma$ ranges in $f$ and $\dot f$ were used to cover all
plausible values.  To
characterize the significance of a putative signal, we used $Z^2_n$
tests with $n=1,2,3$, and $4$.
We repeated the entire search using two reasonable
time steps: 100 days and 365 days.
Finally, to maximize the sensitivity by using
all the data, the power-spectra from each segment were added
incoherently.  No significant pulsed emission is
detected in the 48 searches for each of the five trial ephemerides
that account for the possible aliased X-ray solutions. This outcome 
is not surprising considering the small number of photons, if any,
that can plausibly be attributed to Calvera.

\section{Discussion}

\subsection{Distance and Origin}

The distance to Calvera is poorly constrained by
X-ray spectra, and is uncertain by an order of magnitude.
Upper limits can be derived
from the effective area of thermal X-ray emission, or 
from the conversion of rotational energy to luminosity
in the case that the X-rays are powered by rotation.
Scaling the values of blackbody radius in Table~\ref{tab:spec}
with distance, a radius of $R_1=12$~km for the lower temperature
component in the two-blackbody model corresponds to a distance 
upper limit of 3.6~kpc.   Smaller distances
were derived by \citet{zan11} using hydrogen atmospheres fitted
to the \xmm\ spectra, with $N_{\rm H}$ fixed at the Galactic value,
with the result that $d=1.5-2.3$~kpc.  Here we adopt 2~kpc as an
upper limit on $d$ because, since soft X-ray pulsations
are seen, the star cannot be at uniform temperature and the
effective radius of emission must be smaller than $R_{\rm NS}$.
We adopt 200~pc as a tentative lower limit because the X-ray
column density, which is better measured by the \xmm\
data of \citet{zan11}, is comparable to the total Galactic 21~cm value.

At $d=2$~kpc, the X-ray luminosity of Calvera would be
$L_x=5\times10^{32}$ erg~s$^{-1}$, a typical value for thermal emission
from a NS of its age, $\tau_c=290$~kyr, and consistent
with theoretical NS cooling curves \citep[e.g.,][]{pag09}.
Even at 200~pc, the corresponding $L_x=5\times10^{30}$ erg~s$^{-1}$
is not unusual, since NS temperatures are falling steeply at
this age, particularly those with light-element envelopes.
Even some younger NSs, such as RX J0007.0+7303 in the
supernova remnant CTA~1, are seen to be this faint \citep{hal04},
which could be the effect of a larger mass.
However, the (unmeasured) non-thermal X-ray luminosity of
Calvera {\it would\/} be exceptionally small if at $d=200$~pc, probably
of order $10^{-6}\,\dot E$, while for most pulsars
the ratio $L_x/\dot E$ is greater than $10^{-5}$ \citep{kar12a}.

Calvera stands out as the most extreme case of a ``young'' pulsar
at high Galactic latitude, $b=+37^{\circ}$.  Its X-ray properties
allow it to be nearby, in the Galactic disk, or far enough
to reside in the halo.  If in the halo, it was either born there,
or was ejected from the disk at high velocity,
$v\approx1000\,z_{300}$~km~s$^{-1}$, where $z_{300}$
is its height above the disk in units of 300~pc.  This is near the
upper limit of observed velocities of pulsars \citep{hob05}.
The ejection scenario implies a proper motion
$\mu\simgt\tau_c^{-1}\,{\rm sin}\,b\,{\rm cos}\,b=340$~mas~yr$^{-1}$
to have reached latitude $b=+37^{\circ}$ in $\tau_c=290$~kyr.
To the extent that the true age of a young pulsar is
less than its characteristic age, the proper motion
could be even larger.

If at a distance of $\sim 2$~kpc, Calvera must have been born
in the halo.  If its progenitor was a halo star, or a runaway
O or B star from the disk, then the proper motion of the NS
could be up to a factor of $\sim 3$
smaller than indicated above, and oriented
in any direction after the random supernova kick.
Calvera's proper motion will be measured by an approved
\chandra\ HRC observation in comparison with
the original one from the year 2007 \citep{rut08}.

\subsection{Comparison with \psr}

The pulsar whose circumstances most resemble Calvera's is
\psr, a 0.154~s radio pulsar with
$\tau_c=114$~kyr at $b=+20^{\circ}$ and $d\approx1.24$~kpc,
for which a halo or runaway progenitor scenario has also been suggested
\citep{mcl02}.  The quoted dispersion-measure (DM) distance is from
the Australia Telescope National Facility
catalogue\footnote{http://www.atnf.csiro.au/research/pulsar/psrcat/expert.html}
\citep[version 1.47]{man05}, and is based on the NE2001 Galactic
free electron density model of \citet{cor02}.
The X-ray measured column density to \psr\ \citep{kar12b} is
consistent with the Galactic 21~cm value.
If \psr\ was ejected from the Galactic disk,
its proper motion would have to be $\sim590$~mas~yr$^{-1}$.
However, we can use a pair of existing \chandra\ observations
to rule out this possibility directly.  Inspecting ACIS images
of \psr\ taken in 2001 August (ObsID 1989, 5~ks) and
2010 June (ObsID 11250, 64~ks), its centroid
position is (J2000.0) R.A.=$17^{\rm h}40^{\rm m}25.\!^{\rm s}94$,
decl.=$+10^{\circ}00^{\prime}05.\!^{\prime\prime}9$,
and R.A.=$17^{\rm h}40^{\rm m}25.\!^{\rm s}95$,
decl.=$+10^{\circ}00^{\prime}06.\!^{\prime\prime}1$, respectively.
Within the nominal \chandra\ uncertainty of $0.\!^{\prime\prime}6$,
these positions are indistinguishable from each other
and from the radio timing position,
R.A.=$17^{\rm h}40^{\rm m}25.\!^{\rm s}950(5)$,
decl.=$+10^{\circ}00^{\prime}06.\!^{\prime\prime}3(2)$,
whose epoch is 2000 April \citep{mcl02}.

Taking $0.\!^{\prime\prime}6$ as an upper limit on the displacement
of \psr\ over a 10~yr time span, this implies a
proper motion of $<60$~mas~yr$^{-1}$, or a displacement of
$<2^{\circ}$ since birth, meaning that it was born
outside of the disk if the DM distance is correct.  Note that
this argument is independent of the possible X-ray ``tail'' of
\psr\ \citep{kar08}, whose orientation could be interpreted as
indicating motion parallel to the Galactic plane.   In any case,
the tail may not be real, because the 64~ks \chandra\ image of \psr\
from 2010 doesn't clearly show it.  Curiously, \psr\ is also
undetected in \grs\ by \fermi, as we discuss in Section 4.4.

\subsection{Classification of Calvera}

The place of Calvera among the families of NSs is still not clear
despite our measurement of its spin and spectral properties.
Of the previously hypotheses, we can now rule out only a
mildly recycled pulsar, as Calvera lies far above the spin-up limit
in Figure~\ref{fig:ppdot}.   Its dipole field of $4.4\times10^{11}$~G
is on the low end for ordinary rotation-powered pulsars of the
same characteristic age, but is not unprecedented.
It is difficult to characterize its true (thermal) age using theoretical 
cooling curves, first, because such curves vary,
depending on several unknown parameters.
Second, its X-ray emitting hot spot(s)
do not represent the full surface area of the NS.  Third,
it is entirely possible that some of the thermal emission
is due to polar cap heating by backflowing magnetospheric
particles \citep{har01,har02}.
A typical ratio $L_x/\dot E\sim10^{-3}$ attributed
to this process is seen in thermally emitting millisecond pulsars,
and Calvera's ratio is no larger than this.

Even at the adopted upper-bound distance of $\sim2$~kpc, spin-down power 
could explain all of Calvera's emission.  The pulsed light curves exhibit
a pronounced increase in pulsed fraction with increasing energy
(see Figure~\ref{fig:pulse}), which is characteristic of localized
heating or enhanced conduction from the interior, either of which
are expected at the magnetic poles of a NS.
Preliminary modeling suggests that the pulse profiles as a function
of energy can be fitted assuming two roughly antipodal hot spots differing
in temperature by a factor of two.  A weak, non-thermal spectral
component may be present above 2~keV, as is seen in many radio
pulsars with predominantly thermal spectra, but the photon
statistics of the existing data are not sufficient to distinguish it.

Calvera is of particular interest as the first proposed
candidate for an ``orphaned CCO'' that remains after the host
supernova remnant dissipates \citep{hal11,zan11}.
The three known CCOs pulsars have weak dipole magnetic
fields, in the range $(0.3-1)\times10^{11}$~G, and
negligible spin-down power in comparison with their
bolometric X-ray luminosities of $10^{33}-10^{34}$ erg~s$^{-1}$
\citep{got13}.  An orphaned CCO will remain in the
region of $(P,\dot P)$ space where it was born, and will continue
to cool.  An orphaned CCO could be distinguished from an older pulsar
by its residual thermal X-ray luminosity, which lasts
$10^5-10^6$~yr, and Calvera is of the right age to fill this role.
However, with its larger magnetic field strength and
shorter period than the known CCO pulsars (Figure~1),
Calvera may no longer seem a compelling prototype of a CCO descendant.

Nevertheless, independent of the discovery of Calvera,
a theory for CCOs has been developed that involves burial of a 
typical NS magnetic field ($\sim10^{12}$~G)
by prompt fall-back of a small amount of supernova ejecta,
followed by diffusive regrowth of the same field on a time scale of
$\sim10^4$~yr \citep{ho11,vig12,ber13}.
Magnetic field growth has long been suggested as a reason why
pulsar braking indices are all less than the dipole value of 3. 
In this picture, CCOs would be in a phase of rapid field
growth, with large negative braking index, and their immediate
descendants would lie directly above them on the $(P,\dot P)$ diagram.
The possibility that Calvera, or almost any radio pulsar for that matter,
was once a CCO, cannot therefore be ruled out.  Such a scenario,
in which CCOs quickly join the ordinary radio pulsar population,
carries the added benefit of not requiring yet another
family of NSs to exist that could strain the already evident excess
of pulsars with respect to the Galactic core-collapse supernova
rate, as enumerated by \citet{kea08}. 

\subsection{Absence of \gr\ Emission}

The upper limit that we have derived on Calvera's (assumed isotropic)
$>100$~MeV luminosity, $<3.3\times10^{32}\,d_{2\,\rm kpc}^2$ erg~s$^{-1}$,
is a factor of $\sim200$ below typical pulsars of similar $\dot E$
\citep{abd13}.  \citet{rom11} determined several other \gr\ upper limits
for pulsars from \fermi\ data.  Interestingly, \psr\ mentioned above
is among those not detected, with $L_{\gamma}<5.3\times10^{32}$ erg~s$^{-1}$
as shown in Figure~\ref{fig:fermi}.  This is similar to the limit on
Calvera only if the latter is at its maximum distance of 2~kpc.
Another energetic pulsar of low \gr\ luminosity is PSR J1531$-$5610,
which is weak enough so that only its pulsed luminosity has been measured.
Nevertheless, it is more luminous than Calvera.

If Calvera is one of the nearest pulsars,
at $d\sim200$~pc, then its \gr\ luminosity is $<5.3\times10^{30}$
erg~s$^{-1}$.  This is so far below any other pulsar of comparable
$\dot E$ that it could logically be taken as an argument against
such a small distance.  However, we are not certain what the limits
are on possible beaming corrections for \gr\ pulsars.
In order to explain \gr -weak pulsars such as \psr,
\citet{rom11} favored an interpretation in which an aligned rotator,
with the observer close to the rotation axis,
would direct its outer-gap \gr\ emission away from the observer.
But Calvera is not likely to be an aligned rotator given its large
X-ray pulsed fraction, $\approx30\%$ above 1~keV
(Figure~\ref{fig:pulse}). 
Neither is \psr\ likely to be an aligned
rotator, as its thermal X-ray emission also has a pulsed
fraction of $\approx30\%$ \citep{kar12b}. 
Furthermore, it is difficult to understand the lack
of radio pulsations from Calvera if it is an aligned rotator
unless is it intrinsically radio silent.  The absence of both
radio and \gr\ emission from Calvera must imply a strong constraint
on the beaming patterns and creation of high-energy particles in
pulsars.  Now having two energetic pulsars at high Galactic latitude
with no \gr\ emission to faint limits, we wonder if other undetected
pulsars in the Galactic plane could be similarly weak, while source
confusion and diffuse emission prevents such low upper limits from
being established for them.

\section{Conclusions}

More than three years after 59~ms X-ray pulsations were discovered
from Calvera, we obtained new timing observations with \chandra\
that show its frequency to have changed, corresponding to
$\dot f=(-9.11\pm0.24)\times10^{-13}$~s$^{-2}$.  Interpreting
this as the dipole spin-down of an isolated NS, its
derived properties are $\dot E=6.1\times10^{35}$ erg~s$^{-1}$,
$\tau_c=290$~kyr, and $B_s=4.4\times10^{11}$~G.  These make
Calvera an energetic, young pulsar with a magnetic field toward
the low end of the distribution, and unusual for its
absence of \gr\ emission.

We performed image likelihood and timing searches on almost five
years of \fermi\ data, with a resulting upper limit of 
$<8.2\times10^{-10}$ photons~cm$^{-2}$ s$^{-1}$, or energy flux
$<6.8\times10^{-13}$ erg~cm$^{-2}$~s$^{-1}$ above 100~MeV,
a faint limit that is made possible by Calvera's location at
high Galactic latitude, $+37^{\circ}$.  How extreme this
non-detection is depends on the poorly known distance.
Since X-ray spectral fits have an absorbing column density
greater than or equal to the Galactic value, we estimate a lower
limit on $d$ of $\sim 200$~pc.  A upper limit of $\sim2$~kpc can
be set by requiring the thermal X-ray emitting area, which
dominates the X-ray luminosity, to be less than the surface
area of a 12~km radius NS.  The equivalent upper limit on \gr\
luminosity is $L_{\gamma}<3.3\times10^{32}\,d_{2\,\rm kpc}^2$ erg~s$^{-1}$,
smaller than all \gr\ pulsars of similar $\dot E$.  If Calvera
were as close as 200~pc, the absence of both radio and \gr\
luminosity would be extraordinary, and challenging for
existing models of pulsar acceleration zones
and beaming, especially since Calvera has substantial
X-ray pulse modulation.  The combination of these properties
is difficult to accommodate in an aligned rotator,
or any other geometry.

The radio pulsar \psr\ bears some resemblance to Calvera
in that it is a young, energetic pulsar at high Galactic
latitude that is not a \gr\ source.  We obtained an upper
limit on the proper motion of \psr, $<60$~mas~yr$^{-1}$,
by comparing radio-timing and X-ray positions over a 10 year
baseline.  This tends to rule out that \psr\ was born
in the Galactic disk.  Instead, its progenitor could
have been a runaway O or B star.   We will soon assess Calvera's 
proper motion with an upcoming \chandra\  HRC observation.

It had been speculated that Calvera could be an orphaned CCO,
a descendant of the weakly magnetized X-ray pulsars that fall
unexpectedly in a sparsely occupied region of $(P,\dot P)$ space.
Even though its dipole magnetic field is larger than that of the CCO
pulsars, Calvera could be following the track predicted by
the theory of field burial, with a magnetic field that was
initially submerged by supernova debris, but is rapidly
reemerging and approaching normal strength.  In this theory,
many ordinary radio pulsars may be the orphaned CCOs,
which could be revealed by an excess of thermal X-ray
luminosity with respect to their spin-down ages.

\acknowledgements

This investigation is based on observations obtained with the
\chandra\ X-ray Observatory and \xmm.  \xmm\ is an ESA
science mission with instruments and contributions directly
funded by ESA Member States and NASA.
We have made use of the NASA Astrophysics Data System (ADS)
and software provided by the Chandra X-ray Center (CXC) in the
application package CIAO.  
This publication also makes use of data products from the
Wide-field Infrared Survey Explorer, which is a joint project
of the University of California, Los Angeles, and the
Jet Propulsion Laboratory/California Institute of Technology,
funded by the National Aeronautics and Space Administration.
Financial support was provided
by award GO2-13089X issued by the \chandra\ X-ray Observatory Center,
which is operated by the Smithsonian Astrophysical Observatory for and
on behalf of NASA under contract NAS8-03060.

\end{document}